\documentclass[aps, prb, reprint, superscriptaddress]{revtex4-2}

\usepackage[letterpaper,left=1in,right=1in,top=1in,bottom=1in]{geometry}

\usepackage{amsmath,amssymb,bm}
\usepackage{graphicx}
\usepackage{fourier}
\usepackage{baskervald}

\usepackage{hyperref}
\hypersetup{%
	pdftitle={Deep-Learning-Aided Extraction of Optical Constants in Scanning Near-Field Optical Microscopy}, %
	pdfauthor={Y. Zhao, X. Chen, Z. Yao, M. K. Liu, M. M. Fogler},%
	pdfpagemode={UseNone},%
	pdfstartview={FitH},%
	breaklinks=true,%
	citecolor=blue,%
	colorlinks=true,%
	linkcolor=blue,%
	urlcolor=blue}

\date{\today}

\begin{document}

\title{Deep-Learning-Aided Extraction of Optical Constants in Scanning Near-Field Optical Microscopy}

\author{Y. Zhao}
%\email{yuz289@ucsd.edu}
\affiliation{Department of Physics, University of California San Diego, 9500 Gilman Drive, La Jolla, California 92093, USA}
\author{X. Chen}
\affiliation{Department of Physics and Astronomy, Stony Brook University, Stony Brook, New York 11794, USA}
\author{Z. Yao}
\affiliation{Department of Physics and Astronomy, Stony Brook University, Stony Brook, New York 11794, USA}
\affiliation{Advanced Light Source Division, Lawrence Berkeley National Laboratory, Berkeley, CA 94720, USA}
\author{M. K. Liu}
\affiliation{Department of Physics and Astronomy, Stony Brook University, Stony Brook, New York 11794, USA}
\affiliation{National Synchrotron Light Source II, Brookhaven National Laboratory, Upton, New York 11973, USA}
\author{M. M. Fogler}
%\email{mfogler@ucsd.edu.}
\affiliation{Department of Physics, University of California San Diego, 9500 Gilman Drive, La Jolla, California 92093, USA}

\begin{abstract}

Scanning near-field optical microscopy is one of the most effective techniques for spectroscopy of nanoscale systems. However, inferring optical constants from the measured near-field signal can be challenging because of a complicated and highly nonlinear interaction between the scanned probe and the sample. Conventional fitting methods applied to this problem often suffer from the lack of convergence or require human intervention. Here we develop an alternative approach where the optical parameter extraction is automated by a deep learning network. Compared to its traditional counterparts, our method demonstrates superior accuracy, stability against noise, and computational speed when applied to simulated near-field spectra.

\end{abstract}
\maketitle

\section{Introduction}
\label{sec:1}

\begin{figure}
\includegraphics[width=2.25in]{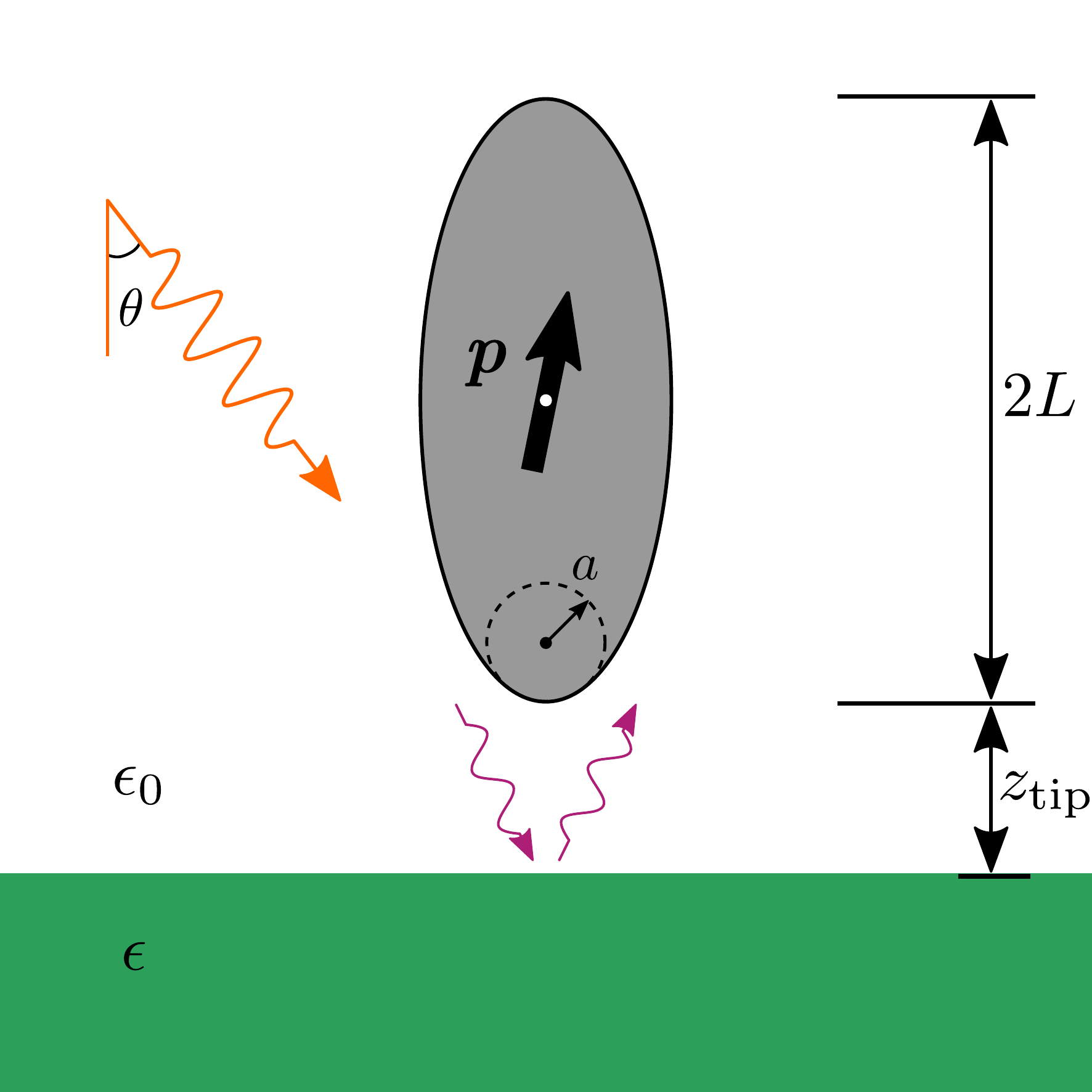}

\caption{A schematic for the theoretical model of an s-SNOM experiment. The probe-sample system is illuminated by an external electric field (orange arrow). Near-field interaction (magenta arrows) between the sample and the probe modifies the dipole moment $\bm{p}$ of the probe, which creates scattered radiation detected in the far field.}
\label{fig:1}
\end{figure}

Scattering-type scanning near-field optical microscopy (s-SNOM) is a powerful method for studying electromagnetic response beyond the diffraction limit \cite{keilmann_near-field_2004, atkin_nano-optical_2012, centrone_infrared_2015}. Recent applications of this technique to two-dimensional materials \cite{fei_gate-tuning_2012, chen_optical_2012, dai_tunable_2014, fei_tunneling_2015}, complex oxides \cite{qazilbash_mott_2007, zhan_metal-insulator_2007, lai_mesoscopic_2010, zhang_nano-resolved_2019}, and organic compounds \cite{nikiforov_probing_2009, westermeier_sub-micron_2014, mastel_nanoscale-resolved_2015, qin_nanoscale_2016} have attracted much attention. In a typical s-SNOM experiment, depicted schematically in Fig.~\ref{fig:1}, a probe with a sharp tip is brought in proximity to a sample of interest. The system is illuminated by a light beam of some frequency $\omega$ and the complex amplitude of the scattered light is measured. While the scattering occurs mostly because of the reflection of the beam from the shank of the probe, the scattering amplitude also contains a correction due to the short-range interaction between the tip and the sample. This correction
is a nonlinear function of the tip-sample distance  $z_{\mathrm{tip}}$.
Therefore, the signal measured in the tip tapping mode contains higher harmonics
of the tapping frequency,
which can be isolated by demodulation with a lock-in amplifier.
This procedure yields a set of complex amplitudes $s^{\bot}_n e^{i{\phi}^{\bot}_n}$ where $n = 2$, $3$, $\dots$ is the demodulation order (see below).

%%
% FIG. 1

A common end goal of s-SNOM experiments is to extract optical parameters of the sample from the near-field spectra, e.g., for a bulk material, to infer its permittivity $\epsilon =\epsilon (\omega)$ from the demodulated amplitudes $s^{\bot }_n(\omega)$ and phases ${\phi }^{\bot }_n(\omega)$. Solving this inverse problem requires modeling of the tip-sample interaction. With a chosen model, $s^{\bot}_n e^{i{\phi}^{\bot}_n}$ is computed using $\epsilon$ as an adjustable parameter, until the calculated spectra agree with the data. A rigorous solution of such a forward problem (f-problem) is computationally expensive, and so the modeling typically involves some approximations. Thus, it has been traditional to represent the probe as a polarizable point dipole \cite{knoll_enhanced_2000}. This simple approach is however not quantitatively reliable. For example, it fails to describe the dependence of the signal on $z_{\mathrm{tip}}$ (the approach curve), which is important in the demodulation procedure. A finite-dipole model \cite{cvitkovic_analytical_2007} gives a more realistic approach curve and approximately accounts for the antenna effect of the probe. Yet it does not correctly reproduce the asymptotic behavior for large $\epsilon $ and scaling with the probe length \cite{jiang_generalized_2016}. Both the point- and the finite-dipole models are quasi-static while in reality retardation also comes into play.

In principle, general-purpose electromagnetic solvers can handle the retardation effects and realistic probe geometry \cite{esteban_simulation_2006, esteban_full_2009, wang_optical_2007, chen_rigorous_2017, mcardle_near-field_2020}. However, data fitting requires finding numerical solutions for large parameter sets ($\omega$, $\epsilon$, $z_{\mathrm{tip}}$), which makes computational cost prohibitive. To address the latter limitation, custom solvers employing the boundary-element method \cite{zhang_near-field_2012, mcleod_model_2014} have been developed, providing a more optimal tradeoff between accuracy and computational resources. On the analytical side, it has been shown \cite{jiang_generalized_2016} that the s-SNOM signal is dominated by a handful of eigenmodes confined in the tip-sample nanogap. These eigenmodes become true resonances of the system when the sample permittivity is equal to certain negative values $\epsilon_k$ that depend on $z_{\mathrm{tip}}$ and the shape and size of the tip. Note that $\epsilon(\omega) = -1$ is the condition for the existence of a surface polariton mode of a planar interface. Therefore, the condition $\epsilon(\omega) = \epsilon_k$ can be interpreted as the criterion for the existence of a surface polariton localized to the tip. Once such polariton eigenmodes are pre-calculated for a given tip geometry, the desired near-field amplitudes can be rapidly computed for any $\epsilon(\omega)$. The calculation of these eigenmodes can be made particularly efficient if the probe shape is approximated by a spheroid \cite{jiang_generalized_2016}.

Even when a fast solver for the f-problem is available, fitting experimental spectra can be challenging because of a nonlinear dependence of the near-field amplitudes on the input parameters. The fitting is commonly done by local optimizers, such as the least squares Levenberg--Marquardt method \cite{more_levenberg-marquardt_1978} or the Trust Region Reflective method (TRM) \cite{branch_subspace_1999}. If the signal contains noise or has a complicated line shape, the convergence of such methods is impeded. One could improve the convergence by regularizing the objective function \cite{ruta_quantitative_2020}, using more sophisticated optimizers, or providing better initial guesses for the iterations. However, because of a nontrivial relation between the s-SNOM spectra and physical parameters [see Fig.~\ref{fig:2}], generating such initial guesses often requires a human effort.

Recently, artificial neural networks (ANNs) have demonstrated excellent capabilities for processing complex data obtained by spectroscopy and imaging, including atomic force microscopy \cite{borodinov_deep_2019, gordon_automated_2020}, Raman scattering spectroscopy \cite{valensise_removing_2020}, scanning tunneling microscopy \cite{lee_deep_2020, gordon_scanning_2019}, scanning transmission electron microscopy \cite{ziatdinov_deep_2017}, and more \cite{wiecha_deep_2021}. The ANN method has been applied to s-SNOM data analysis in Ref. \cite{chen_validity_2021, chen_hybrid_2021}. In those studies, the finite-dipole tip model was used and a mapping between the permittivity $\epsilon$ and the near-field signal was constructed. The mapping was tested on data for several real materials with modestly strong optical resonances such as SrTiO$_3$, SiO$_2$, \textit{etc}. In Ref.~\cite{chen_hybrid_2021}, a so-called hybrid neural network (HNN) solver for the f-problem was developed. The network was trained to reproduce the experimental data using the finite-dipole model prediction for the same $\epsilon$ as an input. Additionally, it was shown that the HNN solver, combined with an advanced local optimizer (differential evolution) can be promising for solving the inverse problem (i-problem), namely, extraction of $\epsilon$ from the experimental data.

In the present work we address the same i-problem but our approach is more streamlined and our ANN, optimizer, and tip model are all quite different.
Although we also use simulated data for the training of our network, we go beyond the finite-dipole approximation. We compute these training data using a more realistic spheroidal-tip model, allowing the tip aspect ratio to be an adjustable parameter. Following \cite{borodinov_deep_2019}, we employ the approach where the ANN provides an initial guess for the TRM fitting. This combines the strengths of the two methods: on one hand, it obviates human intervention and on the other hand, it leverages the fast convergence of the TRM when the initial guess is close to the true ground. We show that our method is highly accurate, robust against noise, and able to handle functions $\epsilon(\omega)$ that contain sharp resonances. To simplify the naming, we refer to the network only part of our implementation as ANN and to the ANN-TRM ``hybrid'' as iHNN where
the ``i'' emphasizes that this is a self-contained solver for the i-problem.

The remainder of the paper is organized as follows. In Sec.~\ref{sec:2}, we provide details on the data generation and the ANN design. In Sec.~\ref{sec:3} we compare the performance of the TRM, ANN, and iHNN. We end with concluding remarks in Sec.~\ref{sec:4}. 

\section{Data Generation and Network Architecture}
\label{sec:2}

\begin{figure}
	\includegraphics[width=2.9in]{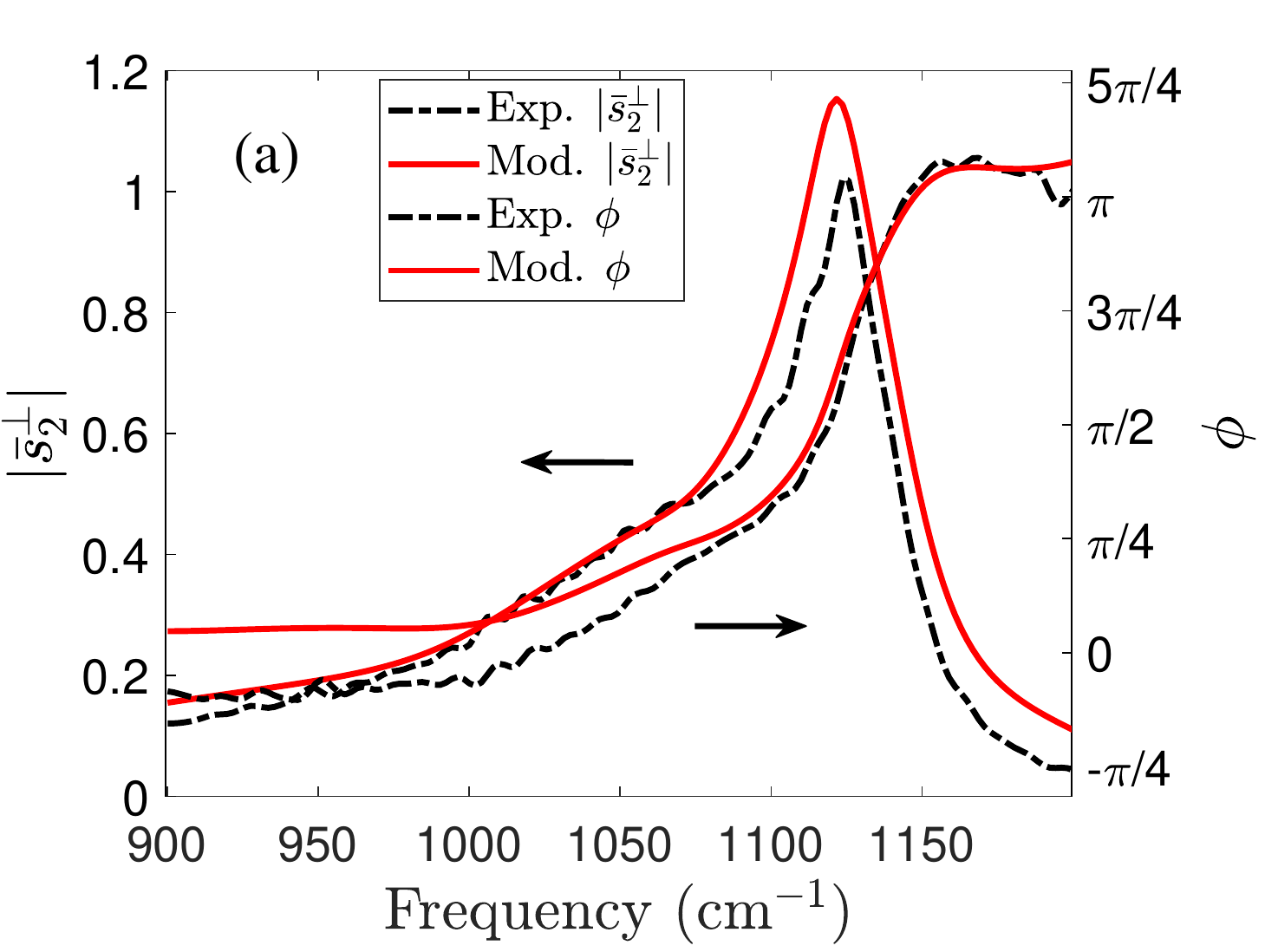}
	\includegraphics[width=2.9in]{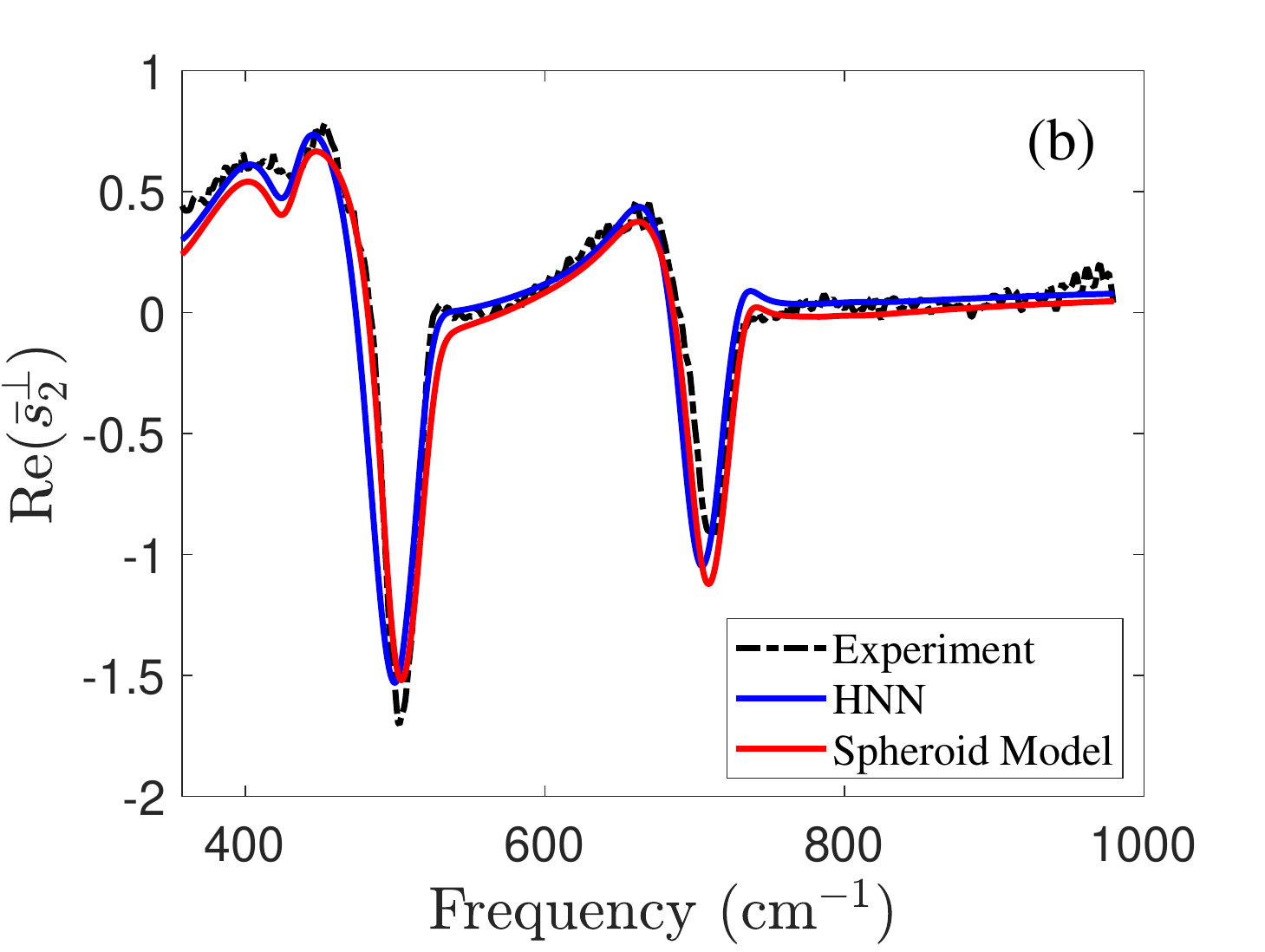}
	\includegraphics[width=2.9in]{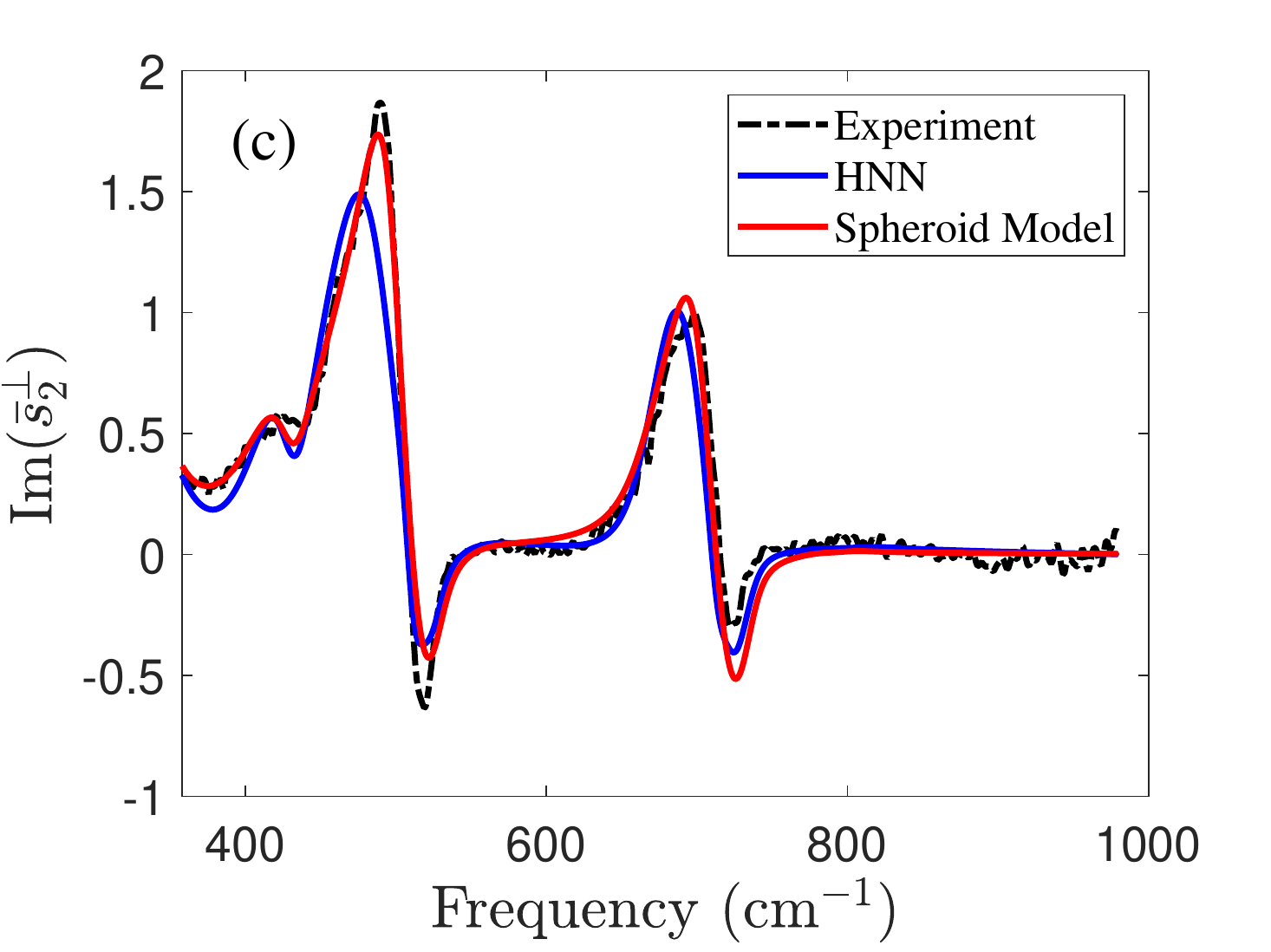}
	
	\caption{Experimental spectra (dashed) and their fits to the spheroidal tip model (solid red) for {(a)} SiO$_2$ {(b, c)} LSAT; the reference material is Au for both. Parameters: {(a)} $z_0 = 0.07a$, $\Delta z = 4.5a$, $L = 5a$, $\epsilon(\omega)$ from Ref.~\cite{kitamura_optical_2007}. {(b)} $z_0 = 0.07a$, $\Delta z = 0.974a$, $L = 5a$, $\epsilon(\omega)$ from Ref.~\cite{nunley_optical_2016}.
	Panels (b,c) include the predictions of the HNN (solid blue) trained on data from panel (a).}
	\label{fig:2}
\end{figure}

To generate large data sets needed for training and testing of the iHNN we use the spheroidal tip model \cite{jiang_generalized_2016}. The major semi-axis of the spheroid has the length $L$ and the radius of curvature of the tip is $a$, see Fig.~\ref{fig:1}. The tip undergoes tapping motion with amplitude $\Delta z$ and frequency $\mathrm{\Omega }$,
\begin{equation} 
 z_{\mathrm{tip}}(\varphi) = z_0 + \Delta z(1 - \cos \varphi),\quad
  \varphi = \Omega t\,,
\label{eqn:EQ__1_}
\end{equation}
where $z_0$ is the minimal tip-sample separation. We compute the observable near-field signal ${\bar{s}}^{\bot }_n(\omega)$ following these five steps. First, we find the near-field reflectivity $\beta $ of the sample-air interface
\begin{equation} 
\beta = \frac{\epsilon(\omega ) - \epsilon_0}{\epsilon(\omega) + \epsilon_0}\ , \label{eqn:EQ__2_} 
\end{equation}
where $\epsilon_0 = 1$ is the permittivity of air. Second, we determine the out-of-plane polarizability $\chi^\bot = \chi^\bot(\beta)$ of the tip as described in Ref. \cite{jiang_generalized_2016}. Third, the polarizability $\chi^\bot$ is demodulated by taking its $n\mathrm{th}$ Fourier transform
\begin{equation} 
{\chi }^{\bot }_n(\omega) = \frac{1}{\pi} \int^{\pi}_0 d\varphi \chi^\bot (\omega, z_{\mathrm{tip}}(\varphi)) {\cos}n\varphi\,.
\label{eqn:EQ__3_}
\end{equation}
(In experiment, the demodulation is done for the far-field background suppression.) Forth, to account for the reflection of the incident and scattered light off the surface, we multiply the result by the so-called far-field factor 
\begin{equation}
F^{\bot }(\omega) = \left(1 + r_{p,\mathrm{FF}}\right)^2 {\sin^2}\theta \,.
\label{eqn:EQ__4_} 
\end{equation}
Here the far-field reflection coefficient $r_{p,\mathrm{FF}}$ is
\begin{equation}
r_{p,\mathrm{FF}} = \frac{\epsilon \cos \theta - \epsilon_0 \sqrt{\frac{\epsilon }{\epsilon_0} - {\sin^2}\theta}}{\epsilon \cos \theta + \epsilon_0 \sqrt{\frac{\epsilon }{\epsilon_0} - {\sin^2}\theta}}\ ,
\label{eqn:EQ__5_}
\end{equation}
and $\theta ={\pi }/{4}$ is the angle of incidence. Finally, the signal is normalized with respect to some reference material ``ref'' (usually, Au or Si):
\begin{equation} 
{\bar{s}}^{\bot }_n\left(\omega \right)=\frac{{\chi }^{\bot }_n\left(\omega \right)F^{\bot }\left(\omega \right)}{{\chi }^{\bot , \mathrm{ref}}_n\left(\omega \right)F^{\bot , \mathrm{ref}}\left(\omega \right)}\ .
\label{eqn:EQ__6_} 
\end{equation}
Although this theoretical model is simplified, it has been shown to provide a good approximation to measured near-field spectra. Figure~\ref{fig:2} demonstrates examples for (a) SiO$_2$ and (b,c) (LaAlO$_3$)$_{0.3}$(Sr$_2$AlTaO$_6$)$_{0.7}$ (LSAT) samples. The best fits to the spheroidal tip model were found via a full search over a grid in the two-dimensional parameter space ($z_0,{L}/{a}$). The material permittivities $\epsilon (\omega )$ used in the calculations were taken from the literature \cite{nunley_optical_2016}. In Fig.~\ref{fig:2}(b,c) we also include the predictions of the HNN~\cite{chen_hybrid_2021} which was first trained on data from Fig.~\ref{fig:2}(a). Even though the fits to the spheroidal tip model and the HNN predictions are both in agreement with the experiment, the former has a twice smaller mean squared error (MSE) and is more accurate near the resonances than the latter. Other examples (see Sec.~\ref{sec:4}) confirm that the spheroidal tip model generally outperforms the finite-dipole one on which the HNN is based.

%%
% FIG. 2
%%

%%
% FIG. 3
\begin{figure*}
\includegraphics[width=4.9in]{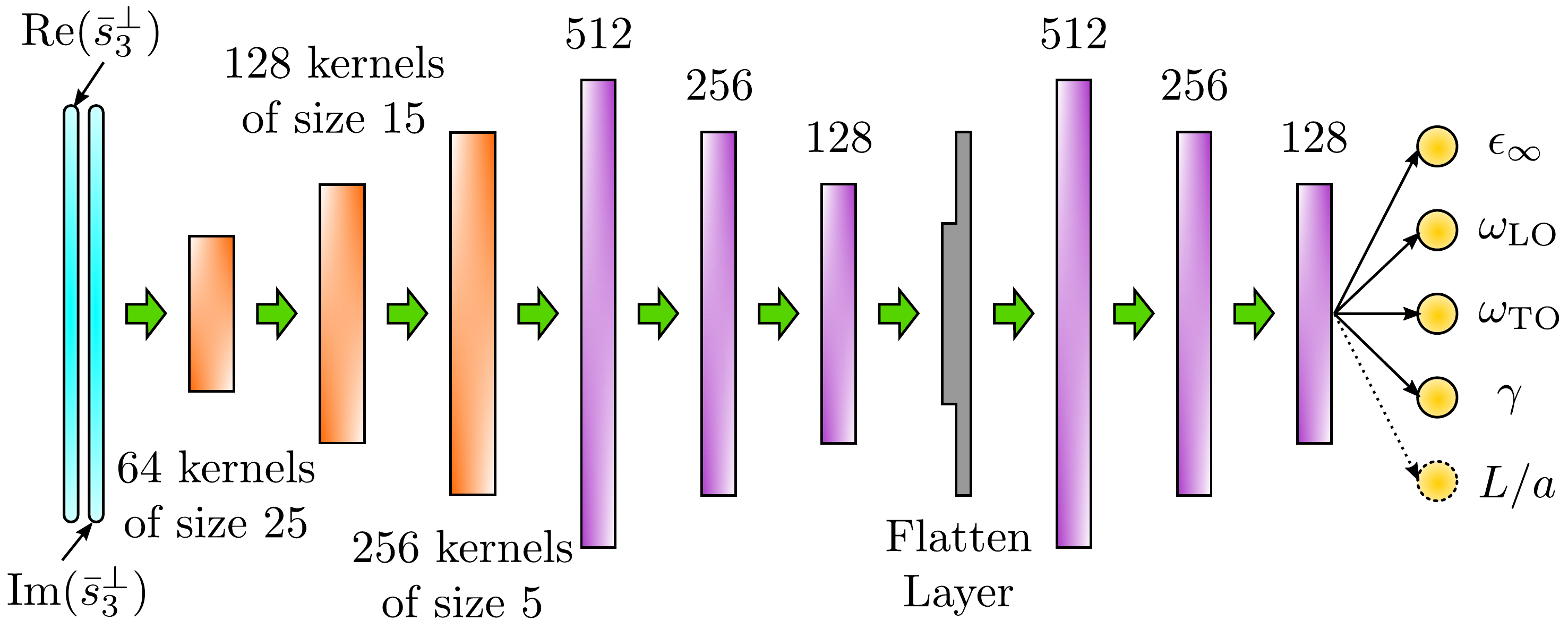}

\caption{The iHNN structure. Complex-valued input signal is separated into its real and imaginary parts. They are sent through three Conv1D Layers (orange) and three Dense Layers (purple) for pattern retrieving. After the Flatten Layer (gray), the recognized patterns go through three Dense Layers (purple) to arrive at the output layer containing the predicted parameters (yellow). The dashed line around ${L}/{a}$ indicates that in the first part of our calculations this parameter was fixed (at $25$) and was not a part of the output.}
\label{fig:3}
\end{figure*}

Fitting of the experimental data often involves several additional adjustable parameters such as $a$ and $\Delta z$. If the frequency-dependent $\epsilon \left(\omega \right)$ is not known, it is an adjustable function as well. As the number of such variables grows, the fitting procedure quickly becomes very laborious. We present an alternative method of data analysis aided by an ANN. For this proof-of-principle demonstration, we consider the dielectric permittivity in the form of a single Lorentzian oscillator:
\begin{equation}
 \epsilon (\omega) = \epsilon_\infty\, \frac{\omega^2_\mathrm{LO} - i\omega \gamma - \omega^2}{\omega^2_\mathrm{TO} - i\omega \gamma - \omega^2}\,,
\label{eqn:EQ__7_}
\end{equation}
where $\epsilon_\infty$ is the high-frequency permittivity, $\omega_\mathrm{LO}$ the longitudinal phonon frequency, $\omega_\mathrm{TO}$ is the transverse phonon frequency, and $\gamma$ is the damping rate. The training data for the ANN were generated as follows. We took Si as our reference material and considered only the $n=3$ demodulation order. We fixed $a=30\,\mathrm{nm}$, $z_0=0.6\,\mathrm{nm}$, and $\Delta z=50\,\mathrm{nm}$. This left us with five free parameters: ${L}/{a}$, $\epsilon_\infty$, $\omega_\mathrm{LO}$, $\omega_\mathrm{TO}$, and $\gamma$. We restricted the first parameter to integer values in the range $5 \le {L}/{a} \le 25$. The last four parameters were drawn randomly from uniform distributions over the intervals $1 \le \epsilon_\infty \le 5$, $1 \le \gamma \le 50$, $650 \le \omega_\mathrm{TO} \le 890$, and $\omega_\mathrm{TO} + 10 \le \omega_\mathrm{LO} \le 900$. (Our frequency unit is $1\,\mathrm{cm}^{-1}$.) These parameter ranges are representative of probes and samples utilized in mid-infrared s-SNOM experiments. Using Eqs.~\eqref{eqn:EQ__1_}--\eqref{eqn:EQ__7_}, we computed ${\bar{s}}^{\bot }_n(\omega)$ spectra, $351$ complex-valued points each, on the uniform (unit spacing) frequency grid $650\le \omega \le 900$. We separated the real and imaginary parts of ${\bar{s}}^{\bot }_n(\omega)$ to form a two-row real-valued matrix. The task of the ANN was to infer the five parameters ${L}/{a}$, $\epsilon_\infty$, $\omega_\mathrm{LO}$, $\omega_\mathrm{TO}$, and $\gamma$ from such input spectra.

Let us describe the structure of the ANN. Its first part is the pattern recognition section comprised of three one-dimensional convolutional layers (Conv1D layers) of sizes $64$, $128$ and $256$, see Fig.~\ref{fig:3}. The layers have kernel sizes of $25$, $15$, and $5$ respectively. The pattern recognition section is completed with three additional dense layers of sizes $512$, $256$, and $128$. The second part of the ANN is the regression section. It begins with a flattening layer where the pattern matrix from the previous section is reshaped into a column vector. This layer is followed by three dense layers of sizes $512$, $256$, and $128$. Finally, the output layer of size $5$ yields the predicted $\epsilon_\infty$, $\omega_\mathrm{TO}$, $\omega_\mathrm{LO}$, $\gamma$, and ${L}/{a}$. In the loss function of the ANN, these output parameters were normalized to be between $0$ and $1$ to help decrease bias during training. By design, Conv1D layers in the pattern recognition section identify patterns of the signal on different frequency scales and the following dense layers analyze these patterns. The regression section then effectively assigns weights to those patterns and combines them to make the predictions. For all layers, the rectified linear unit (ReLU) \cite{nair_rectified_2010} activation is used. We trained this network over 5 epochs utilizing data batches of size 32. The Adam optimizer a constant step size of ${10}^{-4}$ was used for all trainings. The first part of training involved 3,000,000 noiseless spectra generated for the fixed tip semi-length ${L}/{a} = 25$. Subsequently, we added noise and allowed ${L}/{a}=25$ to be a variable parameter. A $20\%$ of these spectra was reserved for validation to combat overfitting. The training took about $4$ hours on a workstation equipped with an NVIDIA $1080$Ti GPU card.

%%
% FIG. 4
\begin{figure}
\includegraphics[width=2.9in]{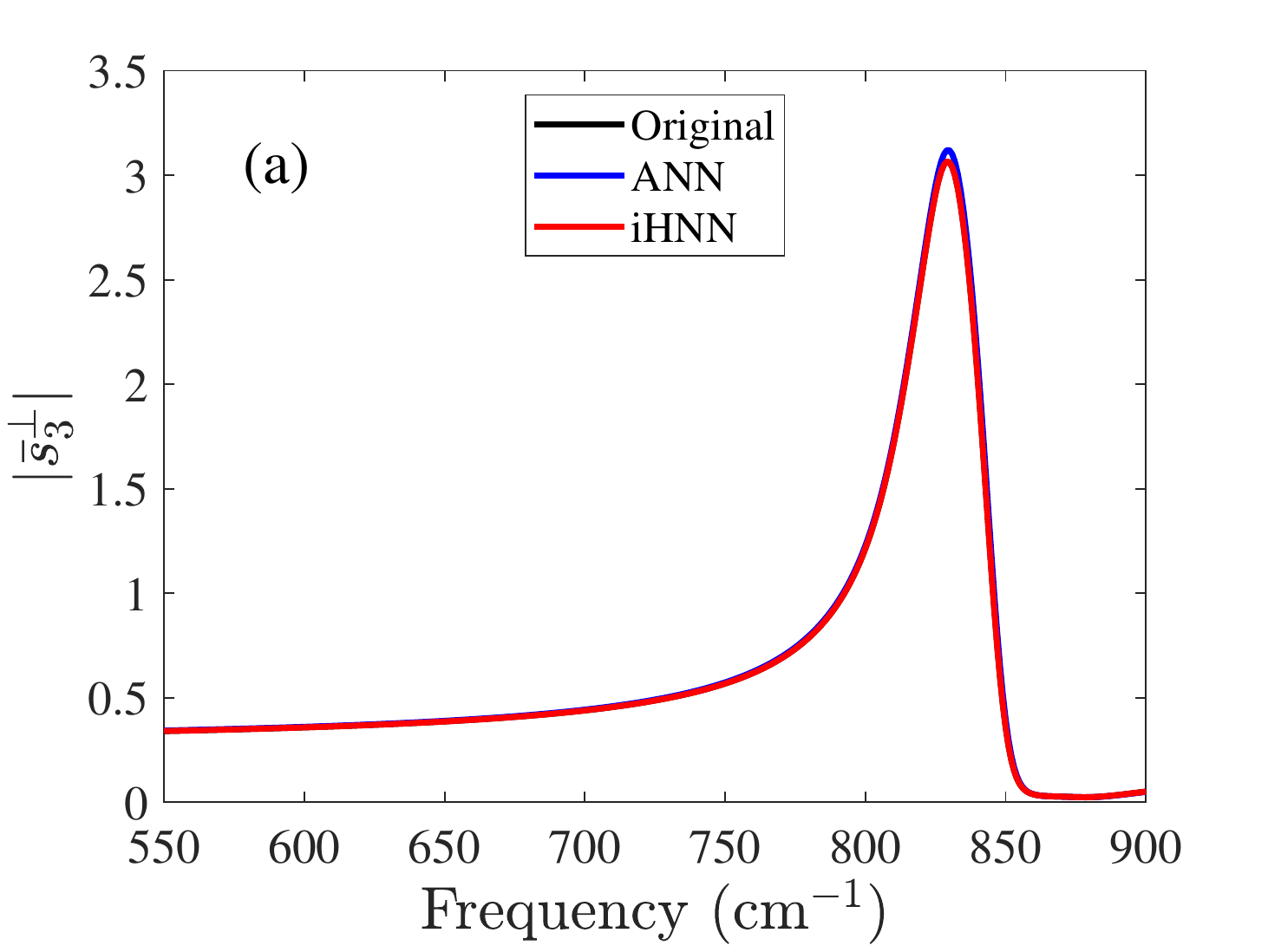}
\includegraphics[width=2.9in]{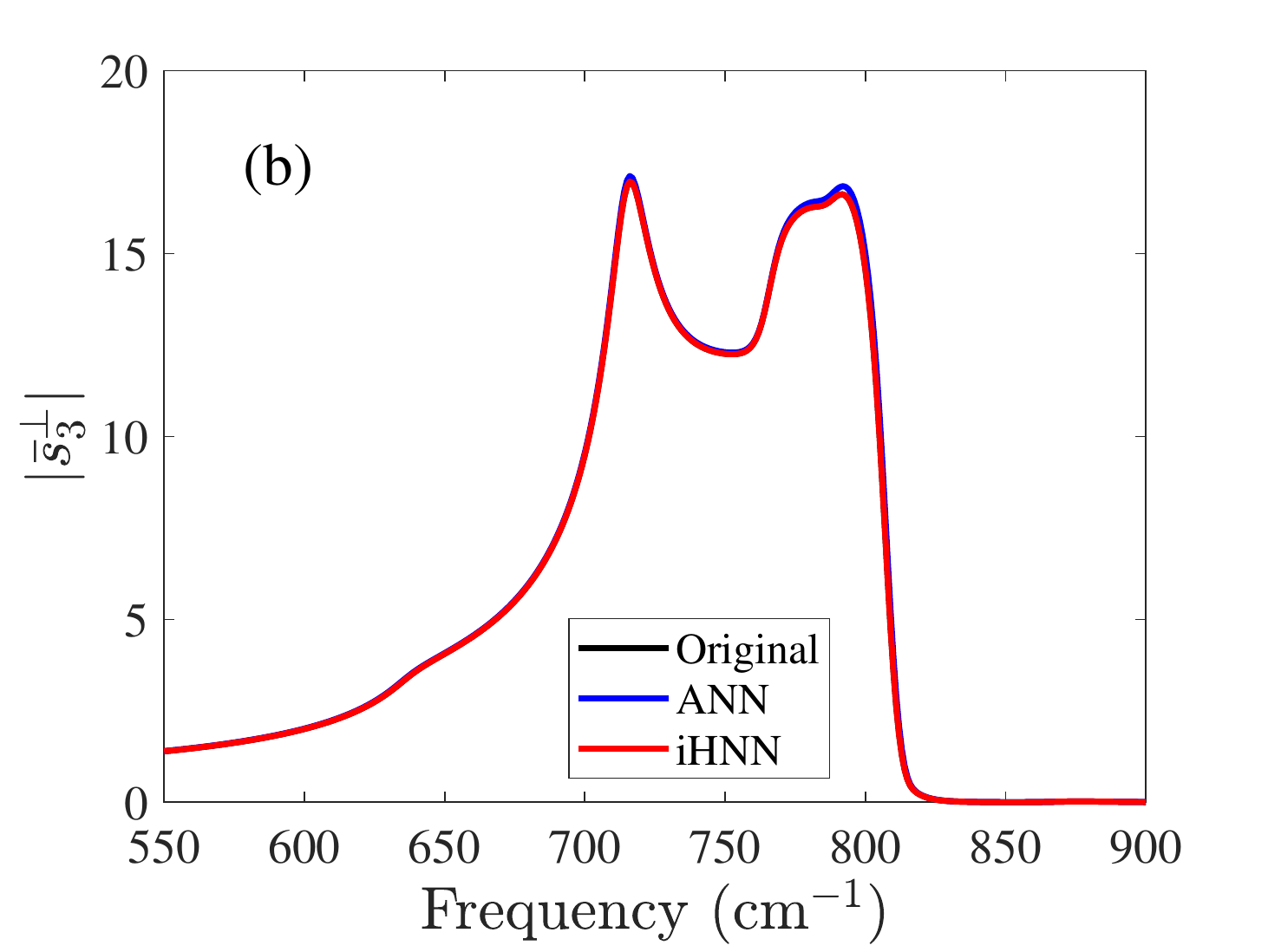}

\caption{A representative near-field spectrum (black) and the spectra computed with the parameters extracted by the ANN (blue) and the iHNN (red).
Due to an almost perfect fit produced by the iHNN, the original data are invisible in the plots.
Parameters: (a) $\epsilon_\infty = 3.79$, $\omega_\mathrm{LO} = 862$,
$\omega_\mathrm{TO} = 814$, $\gamma = 21.8$.
(b) $\epsilon_\infty = 4.36$, $\omega_\mathrm{LO} = 853$,
$\omega_\mathrm{TO} = 636$, $\gamma = 12.9$.
}
\label{fig:4}
\end{figure}

\section{Results}
\label{sec:3}

An example of fitting a noiseless spectrum (with ${L}/{a}=25$) by the ANN is shown in Fig.~\ref{fig:4}. As one can see in Fig.~\ref{fig:4}(a), the fit captures the overall shape of the data well. The difference between the predicted and actual resonant frequency and damping rate are small. Moreover, when the ANN predictions are used as the initial guess for the TRM optimization, the true parameters are found. Note that the spheroidal tip model can generate multiple peaks in the near-field spectra when the damping is sufficiently low~\cite{jiang_generalized_2016}. An example is a double peak seen in Fig.~\ref{fig:4}(b). The ANN is able to handle this case, which demonstrates the capability of this method to analyze complex spectra. 

\begin{figure}
\includegraphics[width=2.9in]{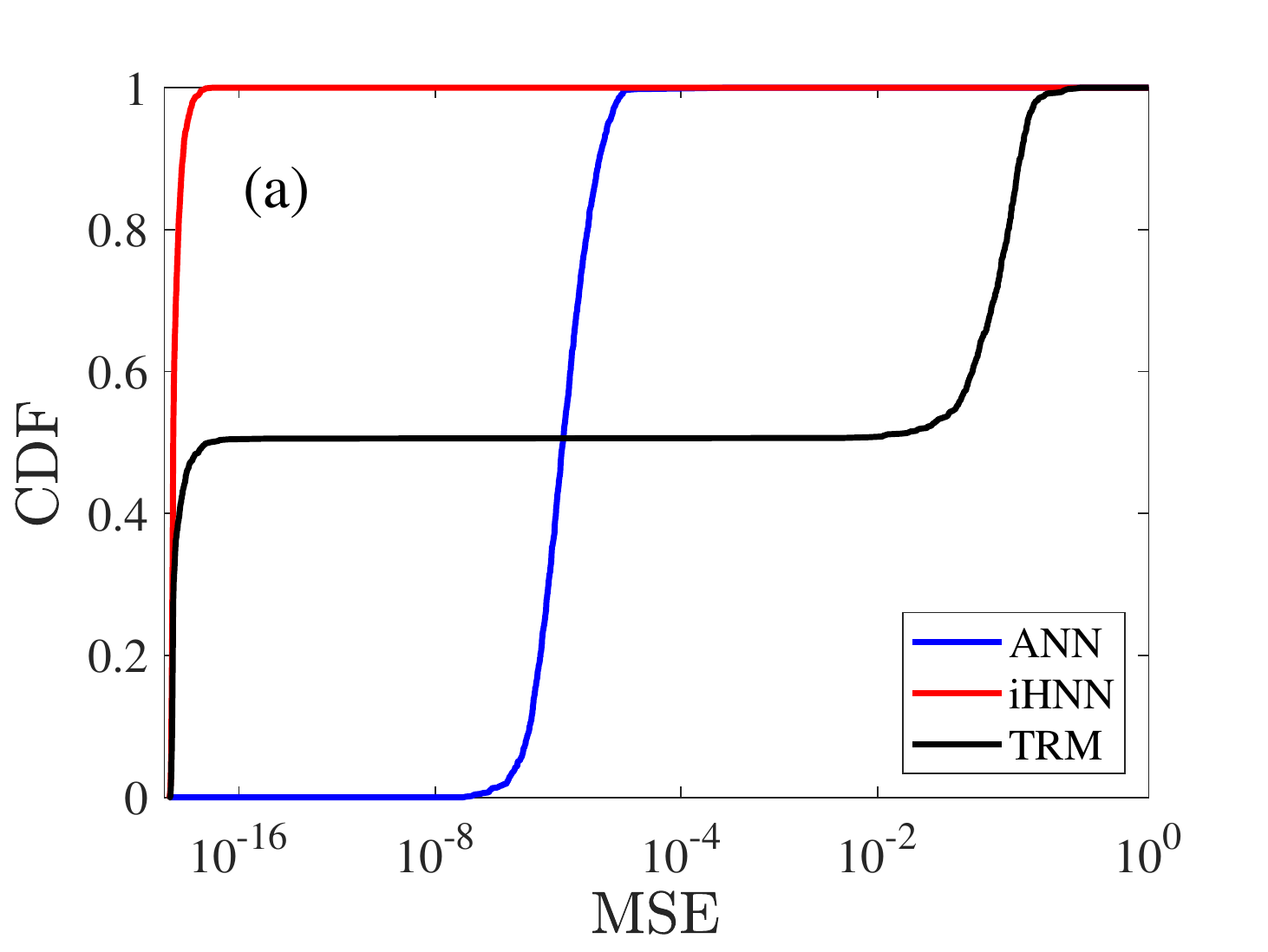}
\includegraphics[width=2.9in]{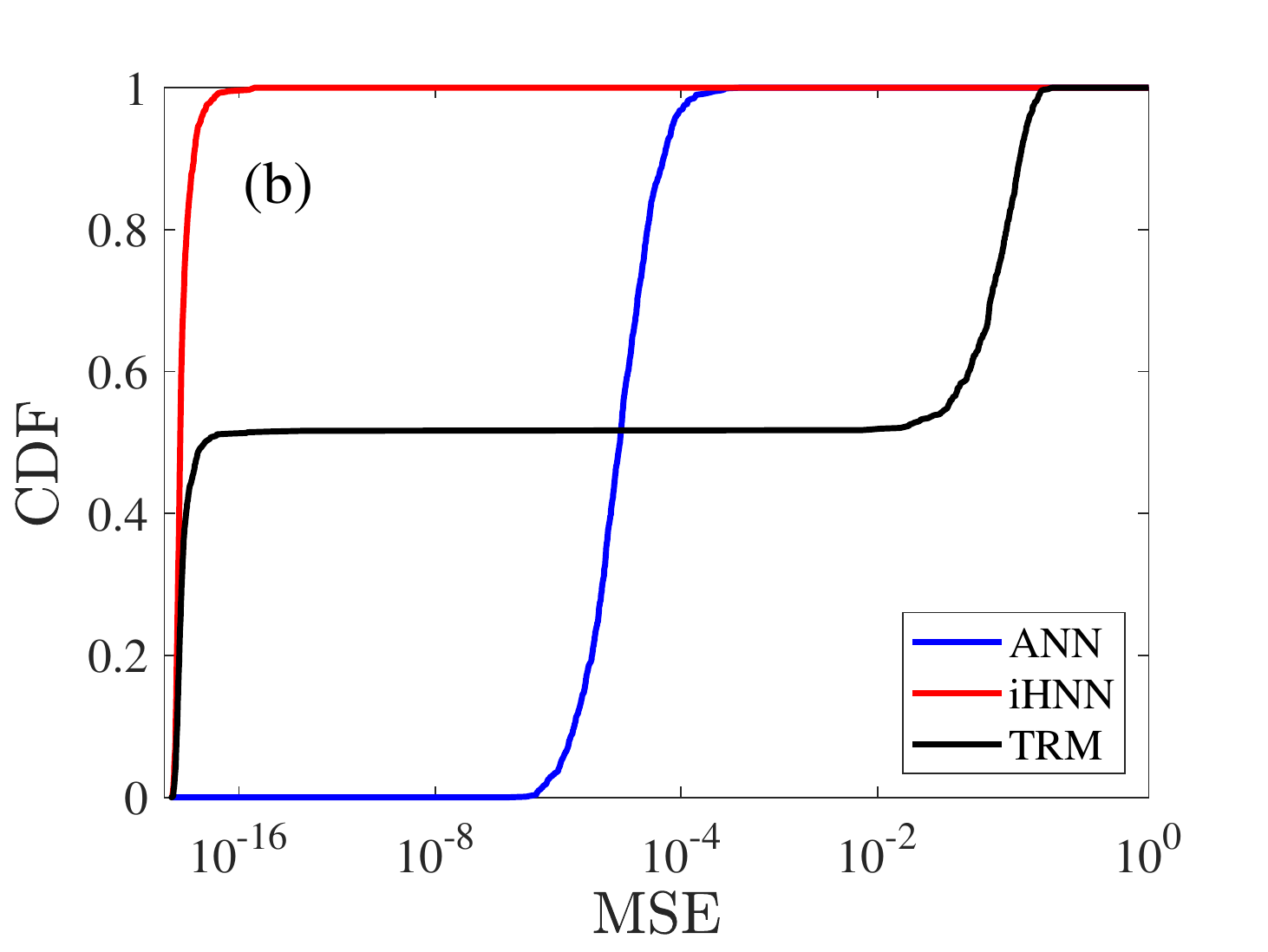}

\caption{Cumulative distribution functions (CDF) of the MSE obtained using TRM (black), ANN (blue), and the iHNN (red), for noiseless spectra. The graphs illustrate that the TRM produces the MSE with the largest mean and standard deviation, followed by the ANN, and then by the iHNN. {(a)} $L = 25a$. {(b)} $5a \le L \le 25a$. 
}
\label{fig:5}
\end{figure}

To test the stability of the TRM, ANN, and their combination iHNN more systematically we used a set of $1,000$ randomly generated spectra. In Fig.~\ref{fig:5}(a) we present the cumulative distribution functions (CDFs) of the MSE of the four parameters $\epsilon_\infty$, $\omega_\mathrm{TO}$, $\omega_\mathrm{LO}$, and $\gamma$. The TRM with a random initial guess (the black curve) gives the largest average and standard deviation for the loss. It is a hit or miss: in roughly half of the cases the TRM was trapped in local minima; however, in the remainder, it reached the global one. In comparison, the ANN alone [the blue curve in Fig.~\ref{fig:5}(a)] provides much more consistent results. The highest MSE of the ANN is around ${10}^{-4}$ and the standard deviation is greatly reduced compared to the TRM. Apparently, the local minima trapping problem is alleviated by the deep connections among neurons. Additionally, for the ANN the average computation time is less than $0.03\,\mathrm{s}$, which is the shortest among the three methods. The iHNN (the red curve) gives the smallest MSE, less than ${10}^{-18}$, effectively reaching the true ground for all spectra. The average computation time for the iHNN remains quite low ($8.2\,\mathrm{s}$). Such results demonstrate that the iHNN combines the advantages of the ANN and the TRM: the former is capable to learn the complex nature of the problem and approximate the required parameters, the latter can swiftly refine these parameter values, while the iHNN can do both.

%%
% FIG. 5
%%

%\begin{table}
%\begin{ruledtabular}
%\begin{tabular}{ccccccc} 
% & \multicolumn{3}{c}{{(a)}} & \multicolumn{3}{c}{{(b)}} \\
% & Original & ANN & iHNN & Original & ANN & iHNN \\ \hline 
%$\epsilon_\infty$ & 3.79 & 3.80 & 3.79 & 4.36 & 4.36 & 4.36 \\
%$\omega_\mathrm{LO}\, ({\mathrm{cm}}^{-1})$ & 862 & 863 & 862 & 853 & 853 & 853 \\
%$\omega_\mathrm{TO}\, ({\mathrm{cm}}^{-1})$ & 814 & 814 & 814 & 636 & 636 & 636 \\ 
%$\gamma\, ({\mathrm{cm}}^{-1}) $ & 21.8 & 21.8 & 21.8 & 12.9 & 12.8 & 12.9 \\ 
%\end{tabular}
%\caption{The original and fitted parameters in Figs.~\ref{fig:4}(a) and (b).
%\label{tbl:1}
%}
%\end{ruledtabular}
%\end{table}

In the spheroidal tip model we use here, the tip geometry is defined by the variables $a$ and $L$ (Fig.~\ref{fig:1}). We added an additional neuron in the output layer of the ANN to make it predict the aspect ratio $L/a$. This five-output-parameter ANN (Fig.~\ref{fig:3}), was trained on $6,000,000$ noiseless spectra generated using $L/a$ ranging from $5$ to $25$ with a unit spacing. After training, the setup was tested on $1,000$ other spectra with ${L}/{a}$ randomly sampled from the same range. In Fig.~\ref{fig:5}(b) we demonstrate the CDF of MSE for the ANN along with those for the TRM and the iHNN. Similar to the fixed $L/a$ case, the TRM converges for only half of the test instances. It also requires the longest computational time, on average $353\,\mathrm{s}$ per run. Despite becoming slightly less precise than our previous four-parameter ANN, the predictions of the new five-parameter ANN are still more accurate and stable compared to the TRM. The ANN predicts the geometric parameter $L/a$ with a mean absolute error of $0.25$ and the standard deviation of $0.34$ for all tested cases. In comparison, the iHNN has arrived at the global minimum for all $1,000$ instances considered. The additional parameter involved in the optimization has increased the average computation time for the iHNN from seconds for the ANN alone to $160\,\mathrm{s}$, although it is still twice faster than the TRM. Based on these tradeoffs between the accuracy and the computational cost, we suggest that the ANN can be used as a high throughput fitting tool for large datasets while the iHNN can fine tune the fits to selected ``interesting'' spectra.

%%
% FIG. 6
\begin{figure}
\includegraphics[width=2.9in]{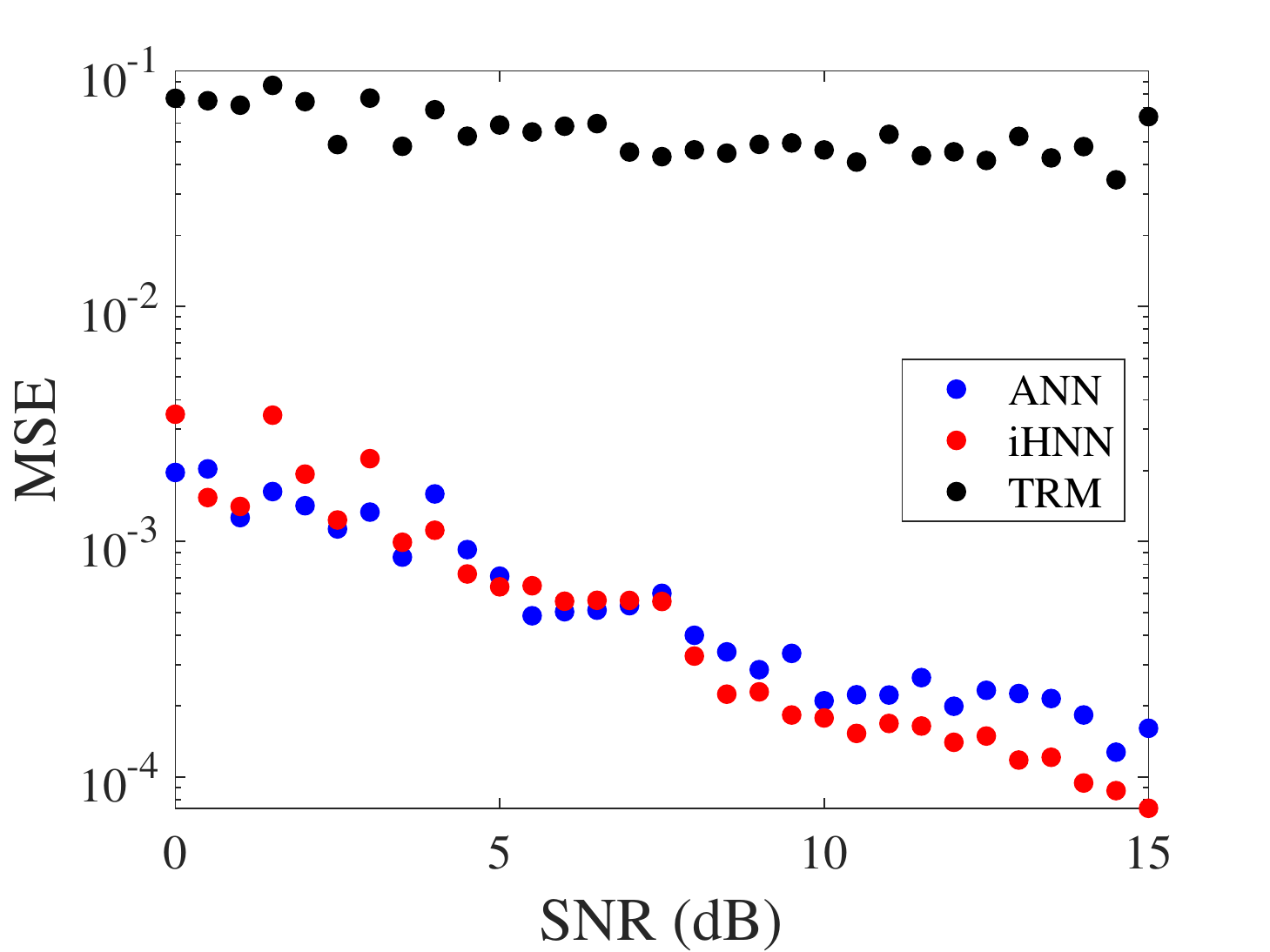}

\caption{MSE obtained from the TRM (black dots), ANN (blue dots), and the iHNN (red dots) as a function of the signal-to-noise ratio (SNR) of the near-field signal.
}
\label{fig:6}
\end{figure}

\begin{figure}
\includegraphics[width=2.9in]{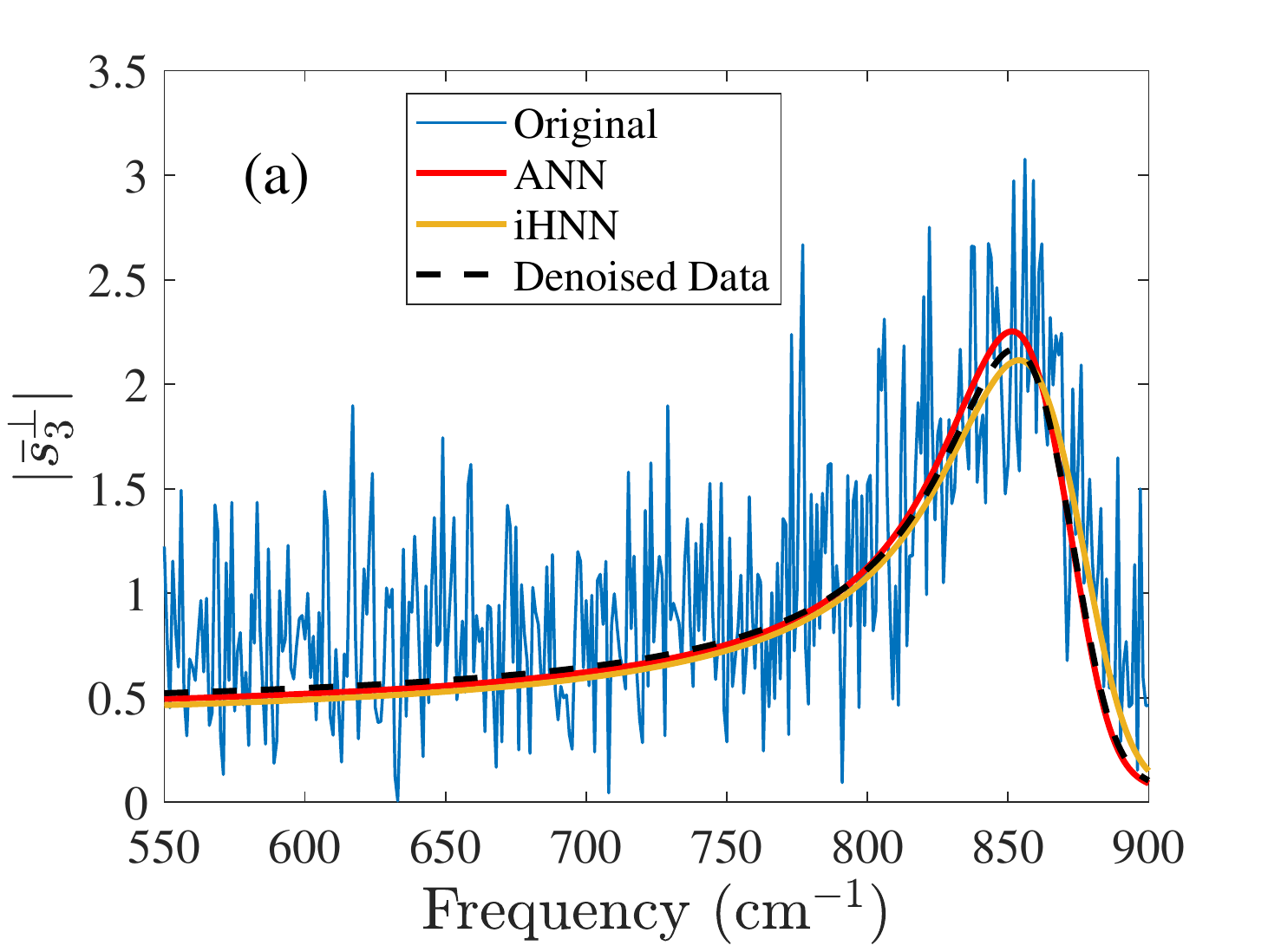}
\includegraphics[width=2.9in]{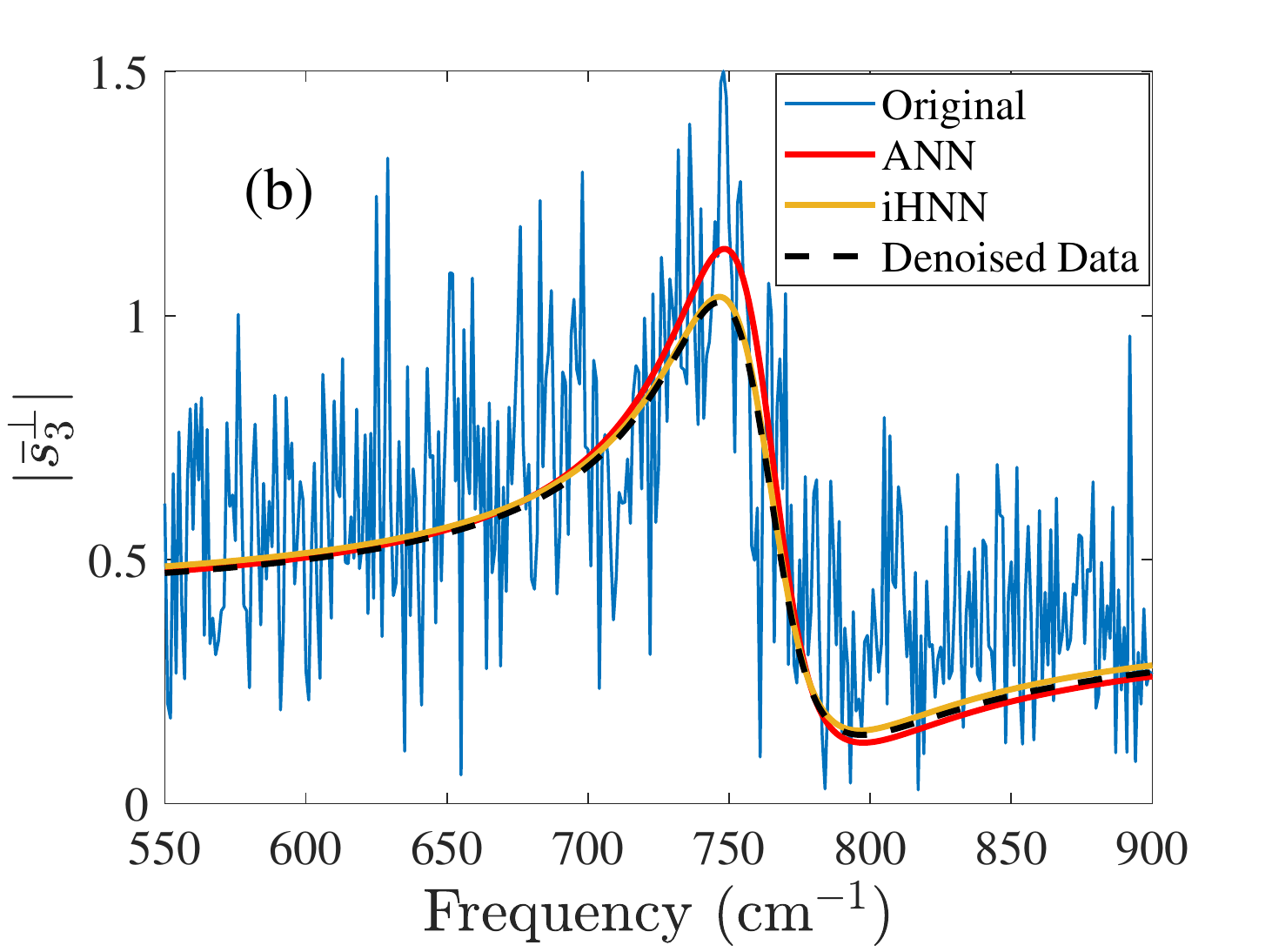}

\caption{Examples of very noisy data (blue) and their fitting by ANN (red) and iHNN (yellow). The dashed line shows the denoised data. {(a)} $\mathrm{SNR} = 6.29\,\mathrm{dB}$ {(b)} $\mathrm{SNR} = 6.77\,\mathrm{dB}$.
}
\label{fig:7}
\end{figure}

Finally, we investigated the effect of a Gaussian noise. We added noise to both real and imaginary parts of the spectra generated for fixed $L/a=25$. The standard deviation for the noise was sampled randomly in the range $[0, 0.7]$ to have a broad range of signal-to-noise ratio ($\mathrm{SNR}$). The final dataset consisted of $6,000,000$ spectra with $\mathrm{SNR} \ge 1$, i.e., nonnegative $\mathrm{SNR}$ on the $\mathrm{dB}$ scale. To focus on highly noisy instances, we further selected $0 \le \mathrm{SNR} \le 15\,\mathrm{dB}$, which is about two thirds of the entire dataset. The same $\mathrm{SNR}$ range was used for generating a test set of $2,000$ spectra. The results are shown in Fig.~\ref{fig:6} where the average MSEs for $\epsilon_\infty$, $\omega_\mathrm{TO}$, $\omega_\mathrm{LO}$, and $\gamma$ are presented. Among the three methods, the TRM performs the worst, giving the lowest average MSE at all $\mathrm{SNR}$. The result hardly improves at higher $\mathrm{SNR}$, which illustrates the impracticability of this method. In contrast, the ANN is more stable and accurate, especially at small SNR. As the SNR gets higher, the average MSE drops linearly and its lead in accuracy increases. Interestingly, the iHNN can be less stable at low $\mathrm{SNR}$s. Under high noise, $\mathrm{SNR} < 5\,\mathrm{dB}$, the MSE of the iHNN can occasionally be more than twice higher than that for the ANN alone. However, at larger SNR, the iHNN gets progressively more accurate than the ANN. We attribute the somewhat reduced performance of the iHNN at small $\mathrm{SNR}$ to the TRM being trapped in one of numerous local minima. An example is shown in Fig.~\ref{fig:7}(a) with $\mathrm{SNR} = 6.29\,\mathrm{dB}$. Despite such an extreme noise level, both the ANN and the iHNN produce good fits. The MSE for the ANN $1.4\times {10}^{-4}$ is actually lower than MSE of $0.0090$ for the iHNN. (The original and predicted values of individual parameters are listed in Table~\ref{tbl:2}.) Note however that the outcome of multi-dimensional optimization depends, in general, on what weights are assigned to different parameters in the total MSE. An example is shown in Fig.~\ref{fig:7}(b). The initial guess by ANN has a more accurate $\epsilon_\infty$. The subsequent optimization by the TRM, which minimizes the unweighted MSE, enlarges the error in $\epsilon_\infty$ but increases the accuracy of $\gamma$, see Table~\ref{tbl:2}.

\begin{table}
\begin{ruledtabular}
\begin{tabular}{ccccccc} 
 & \multicolumn{3}{c}{{(a)}} & \multicolumn{3}{c}{{(b)}} \\
 & Orig. & ANN & iHNN & Orig. & ANN & iHNN \\ \hline 
$\epsilon_\infty$ & 4.91 & 4.65 & 4.34 & 4.92 & 4.89 & 5.05 \\ 
$\omega_\mathrm{LO}$ & 900 & 901 & 909 & 779 & 781 & 779 \\ 
$\omega_\mathrm{TO}$ & 835 & 832 & 834 & 747 & 747 & 748 \\
$\gamma$ & 43.8 & 43.3 & 48.5 & 43.9 & 41.3 & 43.3 \\ 
\end{tabular}
\caption{The original and fitted parameters in Fig.~\ref{fig:7}(a, b).
\label{tbl:2}
}
\end{ruledtabular}
\end{table}

%%
% FIG. 7

\section{Conclusions and outlook}
\label{sec:4}

In our work we address extraction of the optical constants from the near-field signal. In the past, this i-problem has been tackled by conventional optimization algorithms that require an initial guess. A random guess gives about 50\% convergence failure in the examples we studied (Fig.~\ref{fig:6}). We showed that the ANN can provide initial guesses that eliminate such non-converging cases (when no noise is present). Added noise causes it to worsen; however, the prediction accuracy remains at the percent level even for unit signal-to-noise ratio. The advantages of automatic data processing and robustness to noise make our method promising for analyzing actual experimental spectra.

Let us compare the present iHNN with the HNN developed in the earlier work \cite{chen_hybrid_2021}. First, the training data for HNN has been generated from the finite-dipole model whereas the iHNN is based on the spheroidal tip model. The latter tends to handle the highly resonant, i.e., large-$\beta$ [Eq.~\eqref{eqn:EQ__2_}] materials better, see below. Next, the mapping from\textit{ }$\epsilon $ to ${\bar{s}}^{\bot }_n$ by HNN does not utilize the frequency variable. However, due to the nonlinearity of the mapping, there are examples where fairly different $\epsilon $'s can produce very similar ${\bar{s}}^{\bot }_n$'s. These instances make the inverse mapping ambiguous. In contrast, the input given to the iHNN is an entire spectrum, i.e., ${\bar{s}}^{\bot }_n$ as a function of $\omega$. Due to its pattern recognition capability, the iHNN can detect correlation among ${\bar{s}}^{\bot }_n$ data points. This effectively adds hidden labels that help to differentiate ${\bar{s}}^{\bot }_n$ values that can nearly coincide in the pointwise-$\omega$ mapping. Finally, the ``hybrid'' nature of HNN and iHNN is quite different. The HNN uses the output of the theoretical model as an input (basically, an initial guess) whereas the iHNN generates such an initial guess itself and then refines it.

We can think of several directions for improvement and future investigations. One important problem is the lack of universality in the probe-sample coupling. For example, probes made by different manufacturers give somewhat different results for the same sample (see Supplementary Materials, Fig.~S1). Even the signal generated by the same tip changes over time because of rapid wear and degradation. The existing strategy to deal with this variability is to employ models containing adjustable parameters. Such models are either \textit{ad hoc} analytical expressions or true electrodynamic solutions for tips of simplified shapes \cite{knoll_enhanced_2000, cvitkovic_analytical_2007, jiang_generalized_2016, chui_scattering_2018}. The finite-dipole model is of the former kind and the spheroidal tip model \cite{jiang_generalized_2016} used here is of the latter. Both models employ parameter $L$ (Fig.~\ref{fig:1}) which is not the actual length of the tip but an adjustable variable. As shown in Fig.~S3, it is possible to fit the same experimental data for SiO$_2$ and SiC to either model but the corresponding $L$'s differ greatly. The spheroidal tip model appears to give more accurate fits for these materials, cf.~Figs.~S2 and S3. However, the search for a more perfect model remains ongoing.

It is straightforward to upgrade our single Drude-Lorentz resonance approximation for the sample permittivity to a multi-resonance one. This would however increase the number of parameters the ANN has to predict, which may decrease its prediction accuracy. To counteract that, one could explore more advanced ANN architectures, e.g., variational autoencoder \cite{kingma_auto-encoding_2014} or long short-term memory \cite{hochreiter_long_1997}. Finally, we expect a considerable benefit from applying our method to processing near-field spectra from systems that are strongly inhomogeneous or undergoing phase separation \cite{liu_anisotropic_2013, mcleod_nanotextured_2017}. In those systems, spectra collected at different spatial locations can be very dissimilar. The ANN approach offers significant time-saving in analyzing such potentially massive amounts of data.

\acknowledgments

We thank F.~Ruta for useful discussions and A.S.~McLeod for sharing near-field data shown in Fig.~S2. ZY thanks H.~Bechtel, S.~G.~Corder, and M.~Martin for the help in the experiment. The work at UCSD was supported by the Office of the Naval Research under Grant ONR-N000014-18-1-2722. MKL acknowledges support from the NSF Faculty Early Career Development Program under Grant DMR-2045425. This research used resources of beamline $2.4$ at the Advanced Light Source, the Department of Energy Office of Science User Facility under Contract DE-AC02-05CH11231.

\nocite{*}

\bibliography{ANNsSNOM.bib}

\end{document}